\documentclass[prl,aps,amsmath,amssymb,showpacs,twocolumn]{revtex4}
\usepackage{graphicx}
\usepackage{amsmath}
\newcommand{\Boltz}{ k_{\rm\scriptscriptstyle B}}

\newcommand{\Tr}{{\rm Tr}}

\newcommand{\be}{\begin{equation}}
\newcommand{\ee}{\end{equation}}
\newcommand{\bea}{\begin{eqnarray}}
\newcommand{\eea}{\end{eqnarray}}

\newcommand{\quot}[1]{``#1''}

\begin{document}
%\draft
%\preprint{quant-ph xxxxx}
\title{Quantum thermodynamic  Carnot and Otto-like cycles for a two-level system}
\author{ Gian Paolo Beretta }
\affiliation{Universit\`a di Brescia, via Branze 38, 25123
Brescia, Italy} \email{beretta@unibs.it}
\date{\today}
\begin{abstract}From the thermodynamic equilibrium properties of a
two-level system with variable energy-level gap $\Delta$, and a
careful distinction between the Gibbs relation
$dE=T\,dS+(E/\Delta)\,d\Delta$ and the energy balance equation
$dE=\delta Q^\leftarrow - \delta W^\rightarrow$, we infer some
important aspects of the second law of thermodynamics and,
contrary to a recent suggestion based on the analysis of an
Otto-like thermodynamic cycle  between  two values of $\Delta$ of
a spin-1/2 system, we show that a quantum thermodynamic Carnot
cycle, with the celebrated optimal efficiency $1-(T_{\rm
low}/T_{\rm high})$, is possible in principle  with no need of an
infinite number of infinitesimal processes, provided we cycle
smoothly over at least three (in general four) values of $\Delta$,
and we change $\Delta$ not only along the isoentropics, but also
along the isotherms, e.g., by use of the recently suggested
maser-laser tandem technique. We derive general bounds to the
net-work to high-temperature-heat ratio for a Carnot cycle and for
the 'inscribed' Otto-like cycle, and represent these cycles on
useful thermodynamic diagrams.
\end{abstract}


\maketitle

%\narrowtext
\section{Introduction}

Recent studies \cite{Scully,Lloyd,He,Kieu,Kosloff,Feldmann} of
Maxwell demons, quantum heat-engines (often called Carnot engines
even if the cycle is not a Carnot cycle), and quantum heat pump,
refrigeration and cryogenic cycles
 operating between two
heat sources at $T_{\rm high}$ and $T_{\rm low}$  find maximal
efficiencies lower than  the celebrated Carnot net-work to
high-temperature-heat ratio, $1-(T_{\rm low}/T_{\rm high})$. In
particular, Ref. \cite{Kieu}, in studying a specific
two-iso-energy-gap/two-isoentropic-processes Otto-type cycle for a
spin-1/2 system, seems to hint that the quantum nature of the
working substance implies a  fundamental bound  to the
thermodynamic efficiency of heat-to-work conversion, lower than
the celebrated Carnot bound.

Pioneering studies  \cite{Scovil} of  quantum equivalents of the
Carnot cycle for multilevel atomic and spin systems appeared soon
after the association of negative temperatures with inverted
population equilibrium states of pairs of energy levels
\cite{Purcell,Ramsey} and the experimental proof of the maser
principle \cite{Townes}.

In tune with these early studies, here we show  that a Carnot
cycle for a two-level system is possible, at least in principle,
but requires cycling over a range of values of the energy-level
gap $\Delta$. A critical and characteristic feature of this cycle
is that along the isotherms  the value of $\Delta$ must vary
continuously and hence the two-level system must experience
simultaneously a work and a heat interaction. Usually, the
different typical time scales
 underlying mechanical and thermal interactions imply fundamental technological
  difficulties that are among the main reasons
why the Carnot cycle has hardly ever been engineered with normal
substances. In the framework of quantum thermodynamics the
understanding and modeling of mechanical and thermal interactions
is a current research topic, having to do with entanglement,
decoherence \cite{Zurek}, relaxation \cite{Gheorghiu}, adiabatic
(unitary) accessibility \cite{Allahveryan}, but a recent
suggestion by Scully \cite{Scully} indicates that the use of a
\quot{maser-laser tandem} may provide an effective experimental
means to implement the simultaneous heat and work interaction by
smooth continuous change of the magnetic field necessary to
realize the isotherms of our Carnot cycle: the maser serves as the
incoherent ("heat") energy exchange mechanism, the laser as the
coherent ("work") energy exchange.

Consider a  two-level system with a one-parameter Hamiltonian
$H(\Delta)$ such that the energy levels are
$\varepsilon_1=-\Delta/2$ and $\varepsilon_2=\Delta/2$, for
example a spin-1/2 system, in a
 magnetic field of intensity  $B$, with $\Delta=2\mu_B B$ and
$\mu_B=e\hbar/2m_e=9.274\times 10^{-24}\ \rm J/T$  Bohr's magneton
constant.

For our purposes here it suffices to consider the canonical Gibbs
states (the stable equilibrium states of quantum thermodynamics),
i.e., the two-parameter family of density operators
$\rho(T,\Delta)$ with eigenvalues $p$ and $1-p$, mean value of the
energy $E$, and  entropy $S$  given by the relations
\begin{eqnarray}
&&\rho(T,\Delta)=\frac{\exp[-H(\Delta)/\Boltz
T]}{\Tr\exp[-H(\Delta)/\Boltz T]}\ ,\\
&&p=\frac{1}{1+\exp(\Delta/\Boltz T)}=\frac{ 1}{
2}+\frac{E}{\Delta}\ ,\\ &&E=\Tr\rho(T,\Delta)
H(\Delta)=\frac{\Delta}{2}\tanh\left[-\frac{\Delta/2}{\Boltz
T}\right]\!,\ \\ &&S=-\Boltz\Tr\rho\ln\rho=-\Boltz[p\ln
p+(1-p)\ln(1-p)] \ \label{E}
\qquad\nonumber\\&=&-\Boltz\textstyle\left[\left(\frac{1}{2}+\frac{E}{\Delta}\right)\ln
\left(\frac{1}{2}+\frac{E}{\Delta}\right)+\left(\frac{1}{2}-
\frac{E}{\Delta}\right)\ln\left(\frac{1}{2}-
\frac{E}{\Delta}\right)\right]\ .\label{S}
\end{eqnarray}
  We note the following dependences on the
two parameters (temperature $T$ and energy-level gap $\Delta$),
\begin{equation}\label{fundrelation}
p=p(\Delta/T)\ ,\quad E/\Delta=e(\Delta/T)\ ,\quad S=S(\Delta/T) \
,
\end{equation}
and observe that the thermodynamic-equilibrium 'fundamental
relation' $S=S(E,\Delta)$ for this simplest
 system takes the explicit form $S=S(E/\Delta)$ given by the last of  Eqs. ({\ref{S}).
As is well known, all equilibrium properties can be derived from
the fundamental relation.

It is clear from (\ref{fundrelation}) that for an isoentropic
process,
\begin{equation}S = {\rm const\ }
\Leftrightarrow \frac{E}{\Delta}={\rm const\
}\Leftrightarrow\frac{\Delta}{T}={\rm const\
}\Leftrightarrow\frac{E}{T}={\rm const\ }
\end{equation}
and, hence, also the Massieu characteristic function $M=S-(E/T)$
is constant. More generally, it is easy to show that
$dS=(\Delta/T)d(E/\Delta)$ or, equivalently, that the following
Gibbs relation holds for all processes in which the initial and
final states of the two-level system are neighboring thermodynamic
equilibrium states,
\begin{equation}\label{Gibbs}dE=T\,dS + (E/\Delta)\,d\Delta\ .
\end{equation}

Next we write the energy balance equation  assuming that the
system experiences both net heat and work interactions with other
systems in its environment (typically a heat bath or thermal
reservoir at some temperature $T_Q$, and a work sink or source,
respectively),
\begin{equation}\label{Ebalance}dE=\delta Q^\leftarrow -\delta
W^\rightarrow \ ,
\end{equation}
where we adopt the standard notation by which a left (right) arrow
on symbol $\delta Q$ ($\delta W$) means heat (work) received by
(extracted from) the system, when $\delta Q^\leftarrow$ ($\delta
W^\rightarrow$) is positive (negative).

Comparing the right hand sides of Eqs. (\ref{Gibbs}) and
(\ref{Ebalance}), the following identification of addenda,
\begin{eqnarray}\label{heat}
\delta Q^\leftarrow&=&T\,dS \  ,\\ \label{work} \delta
W^\rightarrow&=&(-E/\Delta)\,d\Delta \ ,
\end{eqnarray}
 is tempting and often made and valid, but not granted in general
unless we make further important assumptions. To prove and clarify
this last assertion, we consider two counterexamples, in both of
which the system changes between neighboring thermodynamic
equilibrium states so that both Eqs. (\ref{Gibbs}) and
(\ref{Ebalance}) hold.

As a first counterexample, consider a system which experiences a
work interaction with no heat interaction ($\delta
Q^\leftarrow=0$). The energy change $dE$ is provided by the work
interaction only, while  the entropy change $dS$, required to
maintain the system at thermodynamic equilibrium, is generated
within the system by irreversible relaxation and decoherence
($dS=\delta S_{\rm gen}$). The work is
\begin{equation} \delta
W^\rightarrow=-\frac{E}{\Delta}\,d\Delta-T\,\delta S_{\rm gen}
\quad [\le -\frac{E}{\Delta}\,d\Delta \hbox{ if } T>0],
\end{equation}
and, of course, the process is possible only if $dS\ge 0$.

As a second counterexample, consider a system  which experiences
no (net) work interaction and a heat interaction with a source at
temperature $T_Q$ so that the entropy exchanged with the heat
source is $\delta S^\leftarrow=\delta Q^\leftarrow/T_Q$. In this
case, the energy change $dE$ is provided by the heat interaction
only, while the entropy change $dS$ required to maintain the
system at thermodynamic equilibrium is partly provided by the heat
source and  partly generated within the system by irreversibility
($dS=\delta Q^\leftarrow/T_Q+\delta S_{\rm gen}$). The heat is
\begin{equation}\label{heat2} \delta
Q^\leftarrow=T\,dS+\frac{E}{\Delta}\,d\Delta  \quad [\ne T\,dS
\hbox{ if } d\Delta\ne 0],
\end{equation}
and the process is possible, for $T>0$, only if
\begin{equation}\label{heat3}
\frac{E}{\Delta}\,d\Delta \le \left(1-\frac{T}{T_Q}\right)dE \  .
\end{equation}

It is clear from these two examples, that the correct association
between the work and the heat exchanged, and the energy and
entropy changes, cannot be made by just comparing Eqs.
(\ref{Gibbs}) and (\ref{Ebalance}) without considering also the
entropy balance equation, which specifies unambiguously
 what part of the entropy change is provided by
exchange via heat interaction(s) and what part is generated
spontaneously within the system by its internal dynamics
(relaxation, decoherence). Assuming that the system experiences
both a work interaction and a heat interaction with a heat bath or
thermal reservoir at temperature $T_Q$, the entropy balance
equation is
\begin{equation}\label{Sbalance}dS=\frac{\delta Q^\leftarrow}{T_Q} +\delta
S_{\rm gen}\ , \quad \mbox{with } \delta S_{\rm gen}\ge 0.
\end{equation}
Now, by eliminating $dE$ and $dS$ from Eqs. (\ref{Gibbs}),
(\ref{Ebalance}) and (\ref{Sbalance}), we find
\begin{equation}\label{work2} \delta
W^\rightarrow=-\frac{E}{\Delta}\,d\Delta+\left(1-\frac{T}{T_Q}\right)\delta
Q^\leftarrow-T\,\delta S_{\rm gen} \ ,
\end{equation}
which reduces to Eq. (\ref{work}) if and only if
\begin{equation}\label{sgen}
\delta S_{\rm gen}=\frac{\delta Q^\leftarrow}{T}-\frac{\delta
Q^\leftarrow}{T_Q} \ ,
\end{equation}
i.e., only when entropy generation is due exclusively to the heat
interaction across the finite temperature  difference between  the
system and the heat source, and not to other irreversible
spontaneous processes induced in the system by other interactions
that tend to pull the system off thermodynamic equilibrium.

Eq. (\ref{sgen}) and the condition $\delta S_{\rm gen}\ge 0$ imply
that the system can receive heat only if $-1/T_Q\ge -1/T$, which
for positive temperatures implies $T_Q\ge T$. As  evidenced by
Ramsey \cite{Ramsey},  $-1/T$  measures the thermodynamic
equilibrium escaping tendency of energy by heat interaction, and
is a better indicator of \quot{hotness} than the temperature $T$
because it validly extends to  negative temperature states. Figure
\ref{Figure} shows graphs of energy $E$, entropy $S$, and Massieu
function plotted as functions of $-1/\Boltz T$ and $\Delta$, as
well as graphs of $E$, $S$ and $\Delta$ versus $S$.

\begin{figure*}[t!]
\begin{center}
\includegraphics[width=\textwidth]{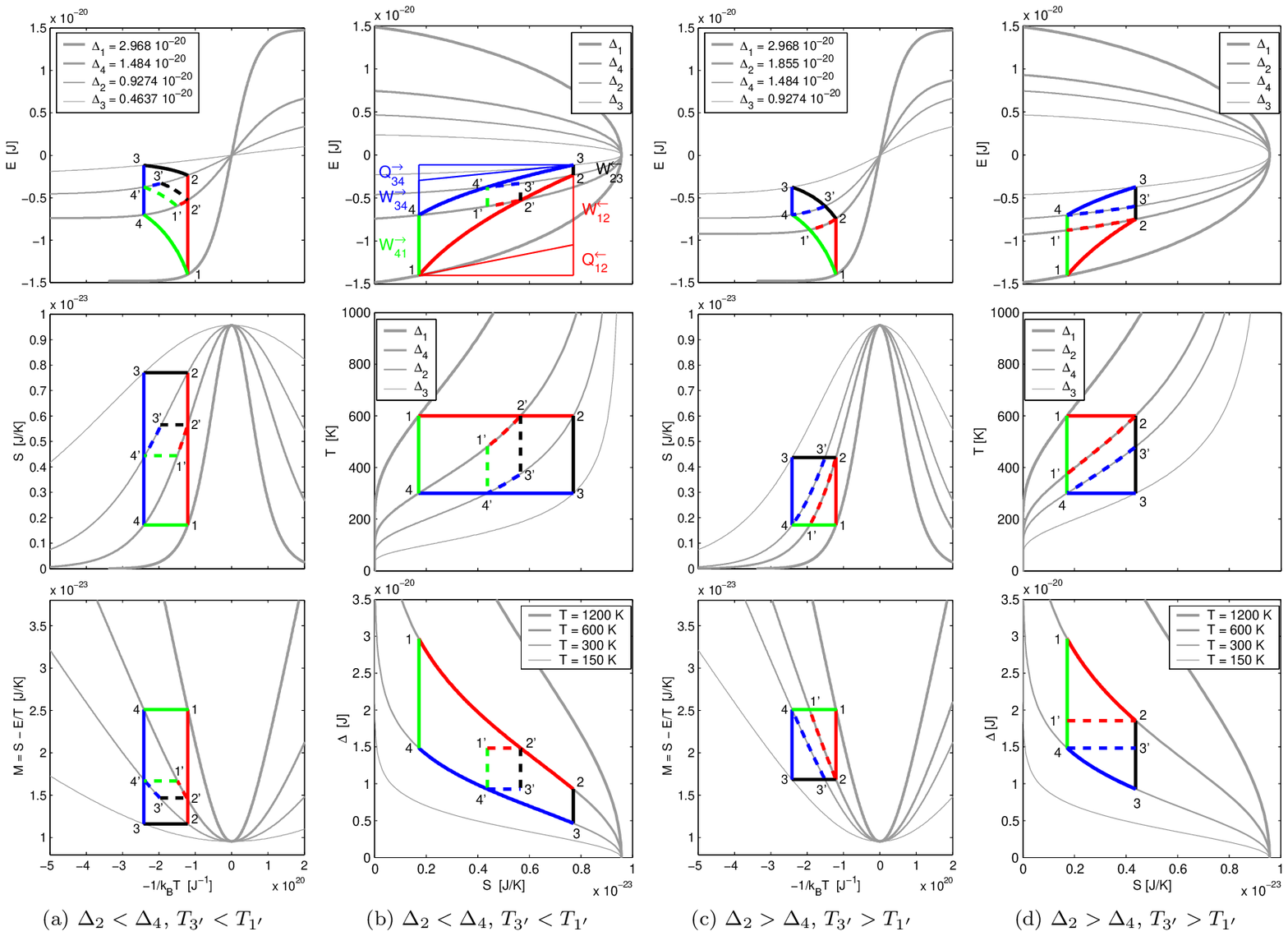}
 \caption{\label{Figure}(Color online) Graphs of thermodynamic equilibrium
properties of a spin-1/2 system. (a) and (c): graphs of energy
$E$, entropy $S$, Massieu function $M=S-(E/T)$ versus $-1/\Boltz
T$ at four values of the energy-level gap $\Delta$. (b) and (d):
graphs of energy
 $E$ and temperature $T$ versus entropy $S$ at four values of $\Delta$,
  and of energy-level gap
$\Delta$ versus $S$
 at four values of temperature.  The four values of $\Delta$ in (a) and (b) correspond to
magnetic field intensities of  1600, 500, 250, and 800 T,
respectively; in (b) and (d) of 1600, 1000, 500, and 800 T (these
extremely large values are chosen for ease of visualization at
ordinary temperatures and intermediate entropies). On each graph,
the 1-2-3-4 paths represents a Carnot cycle: isotherm 1-2 at
$T_{\rm high}=600$ K, isoentropic 2-3, isotherm 3-4 at $T_{\rm
low}=300$ K, isoentropic 4-1. For these cycles,
$Q^\leftarrow_{12}=T_{\rm high}(S_2-S_1)$,
$W^\leftarrow_{12}=E_2-E_1-Q^\leftarrow_{12}$,
$W^\leftarrow_{23}=E_3-E_2$, $Q^\rightarrow_{34}=T_{\rm
low}(S_2-S_1)$, $W^\rightarrow_{34}=E_3-E_4-Q^\rightarrow_{34}$,
$W^\rightarrow_{41}=E_4-E_1$ and, therefore, $W^\rightarrow_{\rm
net,1234}=W^\rightarrow_{34}+W^\rightarrow_{41}-W^\leftarrow_{12}-
W^\leftarrow_{23}=(T_{\rm high}-T_{\rm
low})(S_2-S_1)=Q^\leftarrow_{12}[1-(T_{\rm low}/T_{\rm high})]$.
 The 1'-2'-3'-4' paths [2'=2 and 4'=4 in (c) and (d)] represent
instead an Otto-type cycle of the kind considered in Refs.
\cite{Feldmann,Kieu}: iso-energy-gap process 1'-2' with $T\le
T_{\rm high}=600$ K, isoentropic 2'-3', iso-energy-gap 3'-4' with
$T\ge T_{\rm low}=300$ K, isoentropic 4'-1'. For these Otto-type
cycles, $\Delta_{1'}=\Delta_{2'}=\Delta'_{\rm
high}=\max(\Delta_2,\Delta_4)$,
$\Delta_{3'}=\Delta_{4'}=\Delta'_{\rm
low}=\min(\Delta_2,\Delta_4)$,
$Q^\leftarrow_{1'2'}=E_{2'}-E_{1'}$,
$W^\rightarrow_{4'1'}=E_{4'}-E_{1'}=E_{1'}[(\Delta'_{\rm
low}/\Delta'_{\rm high})-1]$,
$W^\leftarrow_{2'3'}=E_{3'}-E_{2'}=E_{2'}[(\Delta'_{\rm
low}/\Delta'_{\rm high})-1]$ and, therefore, $W^\rightarrow_{\rm
net,1'2'3'4'}=W^\rightarrow_{4'1'}-W^\leftarrow_{2'3'}=
Q^\leftarrow_{1'2'}[1-(\Delta'_{\rm low}/\Delta'_{\rm high})]$. }
\end{center}
\end{figure*}

Each graph in Figure \ref{Figure} shows a Carnot cycle, i.e., a
sequence of an isothermal process 1-2 at a high temperature
$T_{\rm high}$, an isoentropic 2-3, another isothermal process at
a temperature $T_{\rm low}<T_{\rm high}$, and another isoentropic
4-1. Because $S=S(\Delta/T)$ ($S$ decreasing with  $\Delta/T$ for
$T>0$), it is clear that the isoentropic changes between $T_{\rm
low}$ and $T_{\rm high}$
 and the
 isothermal changes between $S_1=S_4$ and $S_2=S_3$ are
possible only by changing the energy-level gap $\Delta$.
Therefore, to have $S_2> S_1$, we need $\Delta_2/T_{\rm
high}=\Delta_3/T_{\rm low}<\Delta_4/T_{\rm low}=\Delta_1/T_{\rm
high}$, i.e.,
\begin{equation}\label{cond2}
\frac{\Delta_{\rm high}}{\Delta_{\rm low}}\left[\frac{T_{\rm
low}}{T_{\rm
high}}\right]^2\!\!=\frac{\Delta_4}{\Delta_2}>\frac{\Delta_4}{\Delta_1}=\frac{T_{\rm
low}}{T_{\rm high}}=\frac{\Delta_3}{\Delta_2}>\frac{\Delta_{\rm
low}}{\Delta_{\rm high}} \ ,
\end{equation}
where, noting that $\Delta_3<\Delta_2<\Delta_1$ and
$\Delta_3<\Delta_4<\Delta_1$, we set $\Delta_{\rm low}=\Delta_3$
and $\Delta_{\rm high}=\Delta_1$. Relations (\ref{cond2}) imply
(see also Figure legend) general bounds on the net-work to
high-temperature-heat ratio (Carnot coefficient),
\begin{equation}\label{bounds}
1-\frac{\Delta_{\rm high}}{\Delta_{\rm low}}\left[\frac{T_{\rm
low}}{T_{\rm high}}\right]^2\!\!<\frac{W^\rightarrow_{\rm
net,1234}}{Q^\leftarrow_{12}}=1-\frac{T_{\rm low}}{T_{\rm
high}}<1-\frac{\Delta_{\rm low}}{\Delta_{\rm high}} \ .
\end{equation}
Notice that $\Delta_2\gtrless\Delta_4$ depending on whether
$\Delta_{\rm low}/\Delta_{\rm high}\gtrless (T_{\rm low}/T_{\rm
high})^2$. Indeed, we may choose arbitrarily $T_{\rm high}$,
$T_{\rm low}<T_{\rm high}$, $\Delta_{\rm high}=\Delta_1$, and
$\Delta_{\rm low}=\Delta_3<\Delta_{\rm high}T_{\rm low}/T_{\rm
high}$. Then, we must set $\Delta_2=\Delta_{\rm low}T_{\rm
high}/T_{\rm low}$ and $\Delta_4=\Delta_{\rm high}T_{\rm
low}/T_{\rm high}$. So, if we choose $\Delta_{\rm low}=\Delta_{\rm
high}(T_{\rm low}/T_{\rm high})^2$, $\Delta_2=\Delta_4$ and we
obtain a Carnot cycle between only three values of $\Delta$.

Of course, if the cycle is reversed, we obtain, instead of a
heat-engine effect, a refrigeration or  heat-pump effect.

Each graph in Figure \ref{Figure} shows also an Otto-type cycle
\cite{Feldmann,Kieu} bound by the same
 $T_{\rm high}$ and $T_{\rm low}$,   i.e., a
sequence of an iso-energy-gap process 1'-2'  at $\Delta'_{\rm
high}$, an isoentropic 2'-3', another iso-energy-gap process 3'-4'
at $\Delta'_{\rm low}$, and another isoentropic 4'-1'. Here, the
fact that $S=S(\Delta/T)$ is a decreasing function of $\Delta/T$
for $T>0$, implies that to have $S_{2'}> S_{1'}$, we need
$\Delta'_{\rm high}/T_{\rm high}=\Delta'_{\rm
low}/T_{3'}<\Delta'_{\rm high}/T_{1'}=\Delta'_{\rm low}/T_{\rm
low}$, i.e.,
\begin{equation}\label{cond3}
\frac{T_{\rm high}}{T_{\rm low}}\left[\frac{\Delta'_{\rm
low}}{\Delta'_{\rm
high}}\right]^2\!\!=\frac{T_{3'}}{T_{1'}}>\frac{T_{3'}}{T_{\rm
high}}=\frac{\Delta'_{\rm low}}{\Delta'_{\rm high}}=\frac{T_{\rm
low}}{T_{1'}}>\frac{T_{\rm low}}{T_{\rm high}} \ .
\end{equation}
Relations (\ref{cond3}) imply (see also Figure legend) general
bounds on the net-work to high-temperature-heat ratio,
\begin{equation}\label{bounds2}
1-\frac{T_{\rm high}}{T_{\rm low}}\left[\frac{\Delta'_{\rm
low}}{\Delta'_{\rm high}}\right]^2\!\!<\frac{W^\rightarrow_{\rm
net,1'2'3'4'}}{Q^\leftarrow_{1'2'}}=1-\frac{\Delta'_{\rm
low}}{\Delta'_{\rm high}}<1-\frac{T_{\rm low}}{T_{\rm high}} \ .
\end{equation}

Notice that $T_{3'}\gtrless T_{1'}$ depending on whether
$(\Delta'_{\rm low}/\Delta'_{\rm high})^2\gtrless T_{\rm
low}/T_{\rm high}$, so that if we choose $(\Delta'_{\rm
low}/\Delta'_{\rm high})^2= T_{\rm low}/T_{\rm high}$,
$\Delta_2=\Delta_4$ and we obtain a special Otto-like cycle with
$T_{3'}= T_{1'}$ and  efficiency $1-(T_{\rm low}/T_{\rm
high})^{1/2}$. Notice also that in terms of the iso-energy-level
gaps of the Carnot cycle that circumscribes the Otto cycle, the
case $T_{3'}> T_{1'}$ obtains for $ (T_{\rm low}/T_{\rm
high})^{5/2}< \Delta_{\rm low}/\Delta_{\rm high} <(T_{\rm
low}/T_{\rm high})^{3/2}$. In this range, the Otto cycle cannot be
run in reverse (refrigeration or heat-pump) mode between two heat
baths, for in such mode the hot bath temperature must be at most
 $T_{1'}$ and the cold bath at least $T_{3'}$.

Because the iso-energy-gap processes (iso-magnetic field for
spin-1/2 system) which characterize the Otto-type cycle are not
isotherms, if they are obtained \cite{Kieu} by contacts with heat
baths at $T_{\rm high}$ and $T_{\rm low}$, respectively, they
involve entropy generation due to irreversibility resulting from
the  heat exchange [see Eq. (\ref{sgen})] across a large
temperature difference (decreasing as $T_{1'}\rightarrow T_{\rm
high}$ and $T_{3'}\rightarrow T_{\rm low}$). These and  other
realistic irreversibilities are modeled in Ref. \cite{Feldmann}
with a Kossakowski-Lindblad-type linear dissipative term in the
quantum dynamical law, as a means to describe relaxation to
equilibrium and decoherence, required for example to decouple the
system from the heat source, i.e., to model dynamically the heat
interactions. The only way to avoid these inefficiencies is the
impractical (infinite) sequence of infinitesimal contacts with
heat baths at different temperatures in the ranges $T_{1'}\div
T_{\rm high}$ and $T_{3'}\div T_{\rm low}$.

In this paper, instead, by showing the feasibility of a Carnot
cycle for a two-level system, with no need of sequences of
infinitesimal heat exchanges with an infinite number of heat
baths, we show that the quantum nature of the working substance
does not impose any fundamental bound, other than the celebrated
Carnot bound, to the thermodynamic efficiency of heat-to-work
conversion when two different temperature thermal reservoirs are
available. The possibility of engineering simultaneously heat and
work interactions as needed for the isotherms of the Carnot cycle
seems within reach of current experiments, e.g., via a maser-laser
tandem technique \cite{Scully}. The Carnot cycle \quot{efficiency}
is higher, as it should, than that of the 'inscribed' Otto-like
cycle at the center of recent studies
\cite{Scully,Lloyd,He,Kieu,Kosloff,Feldmann}.

Only twenty years ago quantum thermodynamics and pioneering
proposals
 to incorporate the second law of thermodynamics into
the quantum level of description were considered
\quot{adventurous} schemes \cite{Nature}.  Discussions in quantum
terms of old thermodynamic problems such as that of \quot{unitary
accessibility} \cite{Allahveryan} or of defining entropy for
non-equilibrium states, were perceived as almost irrelevant
speculations. Today's experimental techniques bring thermodynamics
questions back to the forefront of quantum theory. Remarkably, the
rigorous application of energy and entropy balances, provides
ideas and guidance, and the second law remains a perpetual source
of inspiration towards the discovery of new physics.

\end{document}